%Paper: astro-ph/9512162
%From: Yoshiaki Sofue <sofue@mtk.ioa.s.u-tokyo.ac.jp>
%Date: Thu, 28 Dec 95 12:32:50 JST
%Date (revised): Thu, 28 Dec 95 12:44:39 JST

%1995 Aug submmitted to ApJL, 1995 Dec accepted;
%1995 Dec sent to SISSA,
\magnification=1200
\def\v{\vskip 2mm}
\def\vv{\v\v}
\def\vvv{\v\v\v}

\def\noi{\noindent}
\def\no{\noindent}
\def\section#1{\vskip 8mm \noi {\bf #1} \vskip 8mm}
\def\r{\hangindent=1pc  \noindent}
\def\cen{\centerline}
\def\ce{\centerline}
\def\endpage{\vfil\break}
\def\kms{km s$^{-1}$}
\def\Vlsr{V_{\rm LSR}}
\def\Msun{M_{\odot \hskip-5.2pt \bullet}}

\def\htwo{H$_2$}
\def\twco{${12}$CO}
\def\thco{${13}$CO}
\def\lv{$(l, V)$}
\def\bv{$(b, V)$}

\ce{\bf High-Velocity Molecular Gas in the Galactic Center Radio Lobe}
\vv
\ce{Yoshiaki SOFUE}

\v
\ce{Institute of Astronomy, University of Tokyo,  Mitaka, Tokyo 181, Japan}
\ce{E-mail: sofue@mtk.ioa.s.u-tokyo.ac.jp}

\vvv
\ce{\bf Abstract}  \vv

We point out a possible association of high-velocity molecular gas
with the Galactic Center Radio Lobe (GCL).
A molecular spur in the eastern GCL ridge is receding at
$\Vlsr \sim +100$ \kms, and the western spur approaching at
$\Vlsr \sim -150$ \kms, suggesting a high-velocity rotation of the GCL.
We study the kinematics of the GCL based on these molecular line data.

\v
\ce {Subject headings:  Galaxy: center -- Galaxy: dynamics --
ISM: jet -- ISM: molecules }
\v\v

\section{1. Introduction}

The Galactic Center radio Lobe (GCL) is the most prominent vertical object
among various ejection structures in the nuclear region of the
Milky Way, extending more than 200 pc in an $\Omega$ shape (Fig. 1)
(Sofue and Handa 1984; Sofue 1985).
The lobe ridges comprise  strong vertical magnetic fields of the order
of a milli Gauss (Yusef-Zadeh et al 1984; Tsuboi et al 1986; Sofue et al 1987).
Although of its outstanding appearance, suggesting a high-speed ejection,
no kinematics has been obtained because of the lack of spectroscopic
information.
Recently we (Sofue 1995a, b)
studied three-dimensional structures of the nuclear ring
and expanding features based on the $^{13}$CO $(J=1-0)$ line data from
the Bell-Telephone-Laboratory (BTL) survey (Bally et al 1987),
In the course of data analyses we found high-velocity molecular features
clearly associated with the GCL.
In this Letter we report the discovery of these peculiar features,
and investigate the kinematics of GCL for the first time.

\section{2. Radio Continuum and Molecular-Line Distributions}

A radio continuum map of the galactic center at 10.5 GHz
observed with the Nobeyama 45-m radio telescope at
a resolution of $2'.6$ is shown in Fig. 1 (lower panel) (Sofue 1985).
The GCL comprises the prominent two-horned spurs at $l\sim 12'$
and $l\sim -35'$, extending toward positive latitudes.
The eastern (left-hand side) ridge is an extension from the radio Arc at
$l=12'$, and its negative-latitude extension is also clearly visible.

\ce{-- Fig. 1}

Molecular gas distributions are shown in Fig. 2.
The top panel shows the integrated intensity map of the  \thco\ ($J=1-0$)
line emission  from the BTL survey.
A molecular spur at $l\sim -30'$ is found in positional coincidence
with the western (right-hand side) GCL ridge, extending toward
positive latitudes, and its negative-latitude counterpart is also visible.
However, no negative-$b$ counterpart is present for the $l=10'$ feature.
Another enhanced CO emission is found at $l\sim 10', ~b\sim 10'$ associated
with the eastern GCL ridge.
As is shown later, these eastern and western molecular spurs comprise
high-velocity  gas at $\Vlsr \sim +100$ and $-150$ \kms, respectively.
In order to see how the high-velocity gas is distributed, we
obtained CO intensities integrated from $\Vlsr = -160 $ to $-150$ \kms for
the western half, and from $\Vlsr = +90$ to $+100$ \kms\
for the eastern half.
Result is shown in a composite map in Fig. 2 (bottom), where
the CO  spurs now more clearly show up, resembling the two-horned
continuum ridges.
The eastern CO clump is apparently connected to a broader
feature extending toward $l\sim 20',~b\sim 15'$.
However, this extended feature has an order of magnitude lower intensity,
and can be identified with high-latitude part
of the so called 180-pc expanding ring at  150 to 180 \kms (Sofue 1995a).
Moreover, the GCL clump has a sharp boundary at $l=10'$.
Therefore, the extended feature is more likely to be a distinguished
feature from the GCL molecular clump.
The GCL molecular clumps are also recognized in the total intensity
distribution of the CS line emission from the BTL survey, which
indicates that the GCL ridges contain high-density molecular gas.

\cen{-- Fig. 2 --}

Cross sections of the 10-GHz radio continuum and molecular line intensities
across the GCL ridges at $b=10'$ are compared in Fig. 3.
Since the 10  GHz emission at $b=10'$ is strongly contaminated by the
radio halo of Sgr A, the GCL ridges do not clearly show up.
So, we  present also a 10-GHz cross section at $b=15'$, where the
lobe ridges show up most clearly.
The CO and CS cross sections are almost identical to each other, and both
are  well correlated with the radio ridges.
However, the molecular peaks are slightly shifted inward of the radio peaks
by about $4'$ (10 pc).

\cen{-- Fig. 3 --}% Cross sections, radio/13CO/CS}

\section{3. Kinematics of the High-Velocity Molecular Structures}

Kinematical relation between the molecular and continuum features can be
studied using \lv\ (position-velocity) diagrams at corresponding latitudes.
The upper panel of Fig. 1 shows an \lv\  diagram at $b=8'$ of the CO line.
The eastern and western molecular spurs are recognized as the high-density
clumps at \lv\ $ \sim (8',100$ \kms) and at $(-30', -150$ \kms), respectively.
The two clumps appear about symmetrical with respect to the center of the
GCL, while slightly shifted toward negative velocity by about $- 20$ \kms.
In the same figure we compare this \lv\ plot with the radio continuum map,
where the GCL positions are marked by the thick lines,
and the positional relation between radio and CO features are indicated
by the vertical lines.
The symmetric appearance of the molecular clumps in the \lv\
plot with respect to the GCL center $(l\sim -10')$ suggests
physical connection of the molecular gas with the radio structure.
The fact that the eastern clump is receding and the western clump approaching
us may indicate a rotation of the GCL gas at $\sim 100 - 150$ \kms.
The molecular spurs may be tangential views of a rotating torus or cylinder
of gas in the lower part of the GCL.

Vertical kinematics  along the GCL ridges can be
obtained by \bv\ diagrams at corresponding longitudes.
Fig. 4 shows a composite \bv\ plot of the CO line, where the positive-velocity
side represents a \bv\ plot at $l=10'$ for the eastern ridge,
and the negative-velocity half is for $l=-30'$ for the western ridge.
The major emission at $-10'<b<5'$ in the figure is due to a rotating
molecular ring of larger radius (Sofue 1995a).
The GCL molecular spurs are recognized as two tilted ridges at
$b \sim +5'$ to $15'$ with $\Vlsr \sim +100$ \kms\ and $\sim -150$ \kms,
which appear  symmetric with respect to $\Vlsr=0$ \kms.
The tilt in the \bv\ plot indicates a velocity gradient along the spur
ridge in the sense that $|\Vlsr|$ increases with the height from the
disk plane.

\cen{-- Fig. 4 --}%  bv diagram at l=10', -30'}

\section{4. Discussion}

Mass of the high-velocity  molecular clumps (spurs) can be
estimated from the CO line intensity.
We assume a conversion factor from the \twco\ ($J=1-0$)  intensity  to \htwo\
column density of $\sim 0.92 (\pm 0.2) \times 10^{20}$ \htwo\ cm$^{-2}$/K
\kms\  derived for the Galactic Center (Arimoto et al 1995),
and take the \twco/\thco\ intensity ratio of about 6.2 (Solomon 1979).
{}From Fig. 3 the averaged excess of the $^{13}$CO intensity
within $10'\times 10'$ (25 pc $\times$ 25 pc) region of each molecular clump is
$ 0.06 ~(\pm 0.02)~ {\rm K}~\times 500$ \kms
$\simeq  30~ (\pm 10$) K \kms.
This yields an \htwo\ column density of $1.7~ (\pm 0.6) \times 10^{22}$
\htwo\ cm$^{-2}$.
This corresponds to a density of $\sim 100$ \htwo\ cm$^{-3}$,
if the spurs are tangential views of a cylinder with a line-of-sight depth
of $\sim 50$ pc,
For the Galactic Center distance of 8.5 kpc, we
obtain a total \htwo\ mass of $\sim 3.3~ (\pm 1) \times 10^5 \Msun$ within
the two GCL molecular clumps.
If the observed clumps are tangential parts of a cylinder,
the total mass would be larger.
In fact, we can recognize some excess gas at $Vlsr\sim -50$ \kms\ in Fig. 1,
which may be part of the cylinder.
The kinetic energy of the gas clumps at velocities $\pm 150$ \kms
is of the order of $7 \times 10^{52}$ ergs.
This amount is not surprising, if it is due to the galactic rotation
balancing the gravitational potential of the stellar
mass of $\sim 10^8\Msun$.

Three possible models have been proposed for the origin of the GCL.
(a) Cylindrical outflow model:
This model predicts that the disk gas is vertically
accelerated along poloidal magnetic field, when the field lines are
twisted by accretion of rotating nuclear disk (Uchida et al 1985).
The cylinder structure in this model is consistent with the radio cross section
of the GCL (Sofue 1985).
The outflow velocity is estimated to be of the order of the Alfv\`en
velocity,  and is $\sim 150 $ \kms\ for a gas density of
$\sim 100$ \htwo\ cm$^{-3}$ and magnetic field of a milli Gauss.
In addition, the whole cylinder with gas must be rotating at the disk
rotation speed, which is consistent with the observed high-velocity
rotation in the \lv\ plot comparable to the galactic rotation (Fig. 1).
The systematic velocity shift of $-20$ \kms\ in \lv\ plot would be
accounted for, if the cylinder axis (outflow axis) is inclined by
$\sim 8^circ$ toward the Sun.
The velocity gradient in the $b$ direction (Fig. 4) could be
accounted for, if the magnetic cylinder is conical,
so that the rotation velocity increases with the distance from the disk.
The inward displacement of the molecular gas ridges from
the radio ridges (Fig. 3), as well as the displacement of the cylinder
axis from $l=0'$ remains an open question.
The gas-vs-magnetic field location would depend on the initial
configuration of the poloidal field and accretion disk, and a detailed
modeling is desired (Uchida et al 1985).
(b) Magnetic loop model:
The GCL may be  an $\Omega$ like inflating magnetic loop
anchored to the magnetized nuclear disk (Sofue and Handa 1984).
As the loop is rotating with the disk, the high-velocity
rotation would be also naturally understood.
However, the gas in such a loop may be falling down toward the disk
along the magnetic tube, and therefore, the vertical velocity gradient
would reflect a deceleration of the falling motion near the roots.
(c) Explosion model:
Finally, the GCL may be an expanding shock front associated with an
explosion near the nucleus (Sofue 1984).
The GCL would  be then a dumbbell-shaped gaseous shell accumulated
from the inner region, and therefore, the rotation must be small due to the
angular-momentum  conservation.
The observed high-velocities would be then due to a line-of-sight effect
of localized expanding gas clumps at large azimuthal angles from the node.
The vertical velocity gradient would be due to a faster expansion
into the halo than into the disk.
According to the discussion as above,
we may here conclude that the cylindrical jet model (a)
appears most plausible among the three models.

The author thanks Dr. J. Bally for making him available with the BTL survey
data base.

\section{References}

\parskip=0pt
\def\r{\hangindent=1pc \hangafter=1 \no}

\r Arimoto, N., Sofue, Y., Tsujimoto, T. 1995, submitted AA.

\r{Bally, J., Stark, A.A., Wilson, R.W., and Henkel, C. 1987, ApJ Suppl 65,
13.}

\r{Sofue, Y. 1984, PASJ 36, 539.}%Halo shocked shell

\r Sofue, Y. 1985 PASJ  37, 697.

\r Sofue, Y. 1995a, PASJ submitted.

\r Sofue. Y. 1995b, PASJ submitted.

\r{Sofue, Y., and Handa, T. 1984, Nature 310, 568.}

\r Sofue, Y.  Reich, W., Inoue, M., Seiradakis, J. H. 1987 PASJ 39, 95.

%\r{Handa, T., Sofue, Y., Nakai, N. Inoue, M., and Hirabayashi, H. 1987, PASJ
%%39, 709.}

\r{Solomon, P.M., Scoville, N.Z., and Sanders, D.B., 1979, ApJ  232, L89.}

\r Tsuboi, M., Inoue, M., Handa, T., Tabara, H.,  Kato, T., Sofue, Y.,
Kaifu, N.1986 AJ 92, 818.

\r Uchida, Y., Shibata, K., Sofue, Y. 1985 Nature 317, 699.

\r Yusef-Zadeh, F., Morris, M., Chance, D. 1984, {Nature}, { 310}, 557.

\endpage
\section{Figure Captions}
\parskip 0pt

Fig. 1: 10.5 GHz radio continuum map of the Galactic Center (lower panel),
and a \thco\ \lv\ diagram at $b=8'$ (top panel).
High-velocity CO clumps are associated with the ridges
of the Galactic Center radio Lobe.
CO and radio contours are at 0.2 and 0.03 K $\times$ 1, 2, .. 10, 12, ..,
20, 25, .., 40, respectively.

\vv
Fig. 2: [Top panel] \thco\ total intensity map (Bally et al 1987).
[Bottom] Composite CO intensity map integrated from $-160$  to $-150$ \kms
for the eastern half, and  from +90 to +100 \kms for the western half.
Contours are at 0.1 K $\times$ 1, 2, .., 10, 12, 15, 20, .., 40, 50, .., 80.

\vv Fig. 3: Cross sections of the radio continuum at $b=10'$ and 15',
CO- and CS-line intensities at $b=10'$.
Note the intensity scale of the 10-GHz emission at $b=10'$ is 1.6 times
that at $15'$.

\vv Fig. 4:  Composite \bv\ diagram of the CO emission.
The upper half ($\Vlsr>0$) is along the eastern ridge at $l=10'$,
and the lower half ($\Vlsr<0$) along the western ridge at $l=-30'$.
Contours are at 0.1 K $\times$ 1, 2., .., 10, 12, 15, 20, .., 40.

\bye